\documentclass[5p, twocolumn]{elsarticle}

\usepackage[LGRgreek]{mathastext}
\usepackage[percent]{overpic}
\usepackage{upgreek}
\usepackage{caption}
\usepackage{subcaption}
\usepackage{float}

\journal{Solid State Electronics}

\begin{document}

\begin{frontmatter}

\title{An Atomistic Modelling Framework for Valence Change Memory Cells}

\author[1]{Manasa Kaniselvan}
\author[1]{Mathieu Luisier}
\author[1]{Marko Mladenovi\'c \corref{mycorrespondingauthor}}
\address[1]{Integrated Systems Laboratory, ETH
Zurich, CH-8092 Zurich, Switzerland}

\cortext[mycorrespondingauthor]{Corresponding author}
\ead{mmladenovic@iis.ee.ethz.ch}

\begin{abstract}
We present a framework dedicated to modelling the resistive switching operation of Valence Change Memory (VCM) cells. The method combines an atomistic description of the device structure, a Kinetic Monte Carlo (KMC) model for the creation and diffusion of oxygen vacancies in the central oxide under an external field, and an \textit{ab initio} quantum transport method to calculate electrical current and conductance. As such, it reproduces a realistically stochastic device operation and its impact on the resulting conductance. We demonstrate this framework by simulating a switching cycle for a TiN/HfO$_2$/TiN VCM cell, and see a clear current hysteresis between high/low resistance states, with a conductance ratio of one order of magnitude. Additionally, we observe that the changes in conductance originate from the creation and recombination of vacancies near the active electrode, effectively modulating a tunnelling gap for the current. This framework can be used to further investigate the mechanisms behind resistive switching at an atomistic scale and optimize VCM material stacks and geometries.
\end{abstract}


\end{frontmatter}

%\linenumbers

%***********************
% Introduction
%***********************

\section{Introduction}

Neuromorphic computing units require the development of solid-state synapses that are often realized in the form of devices with adjustable electrical conductance. Amongst such devices are valence change memory (VCM) cells. These are two-terminal metal-oxide-metal stacks across which applied voltages can drive the creation of oxygen vacancies and their redistribution into a conductive filament - an applied voltage of the opposite polarity can then partially dissolve the filament, thus breaking the conductive pathway across the oxide. The resulting conductance can then be measured with a low readout voltage without further disrupting the defect arrangement. VCMs are both relatively straightforward to fabricate and exhibit an especially large dynamic range. 

Simulation of these devices is complicated by the stochasticity of their operation and the atomistic granularity of the resulting conductance. The movement of oxygen vacancies has been modelled using Kinetic Monte Carlo (KMC) methods \cite{Padovani2017, Zeumault2021} or Molecular Dynamics (MD) simulations \cite{Urquiza2021}, and the microscopic nature of bonding across the defective interfaces of a VCM filament has been investigated at the \textit{ab initio} level \cite{Padilha2018}. Previous simulation frameworks which attempt to describe the full device-level operation of VCM typically combine these methods with trap-assisted tunneling models of current flow \cite{Padovani2017, Bersuker2011, Kopperberg2021}. However, the required equations are derived for single trap energies, and may not fully describe transport through the inhomogeneous defect distribution in a VCM. Capturing both atomic rearrangement and realistic transport properties necessitates a more fundamental level of theory. 
 
\begin{figure}
    \includegraphics[width=\columnwidth]{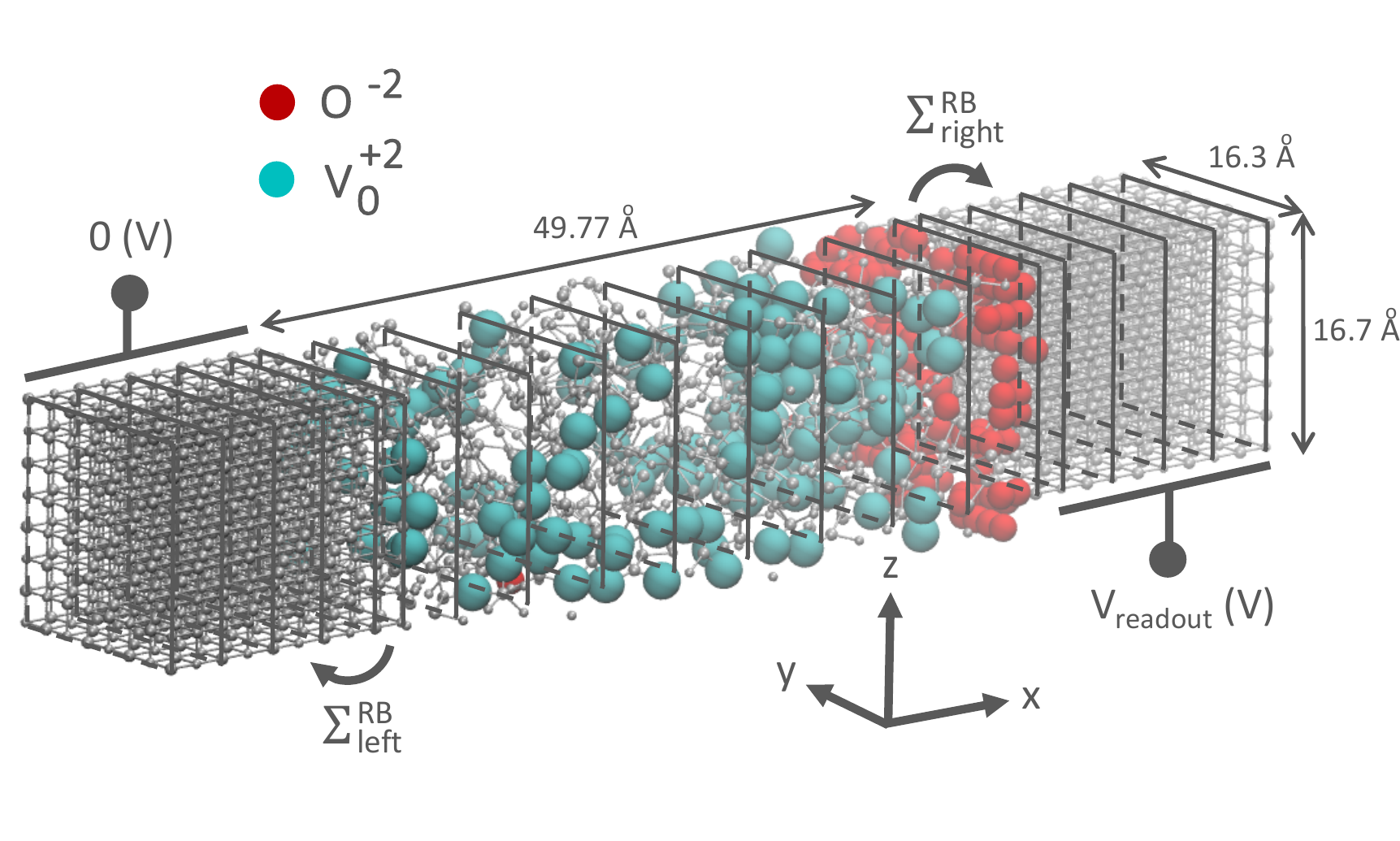}
     \caption{Atomic structure of a typical metal-oxide-metal VCM cell. Oxygen vacancies (V$_0^{+2}$) and oxygen ions (O$^{-2}$) are pictured with larger blue and red spheres, respectively. The device is partitioned into 'blocks' of atoms representing the underlying block-structure of the Hamiltonian ($H_{cp2k}$) and overlap ($S_{cp2k}$) matrices in \textbf{Eq. (2)}.}
\end{figure}

\begin{figure*}[t]
    \includegraphics[width=\textwidth]{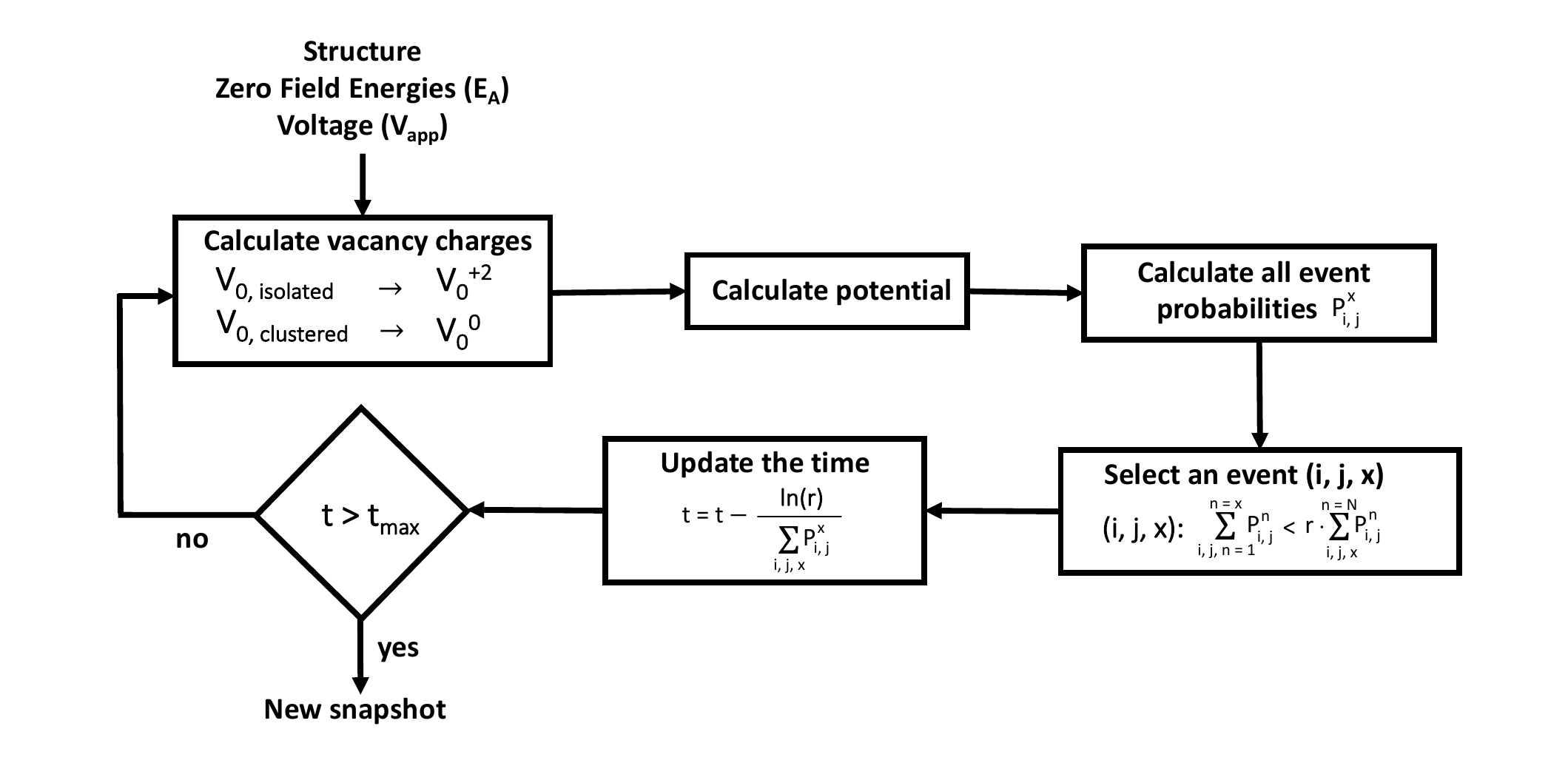}
     \caption{Flowchart of the developed Kinetic Monte Carlo model which determines the state of atomic rearrangement under $V_{app}$, where $r$ represents a random number. The final snapshot of the device is generated at time $t_{max}$. $P^x_{i,j}$ is calculated according to \textbf{Eq. (1)}.}
\end{figure*}

Here we present a framework dedicated to modelling VCM. Our method relies on a stochastic Kinetic Monte Carlo (KMC) model parameterized with density functional theory (DFT), followed by \textit{ab initio} quantum transport simulations, all of which are preformed on the same atomic grid. It thus captures the growth and dissolution of oxygen vacancy filaments through the VCM cell and the electrical current that flows through them.

%***********************
% Methods
%***********************

\section{Modelling Framework}

\textbf{Figure 1} shows a schematic of the nominal VCM cell considered: an amorphous HfO$_2$ oxide with TiN electrodes. The amorphous oxide is initially generated by melting and cooling it from the monoclinic phase using MD. In this process a block of monoclinic HfO$_2$ is first melted at 3000 K for 300 ps, then cooled to 300 K at a rate of 9 K/ps, and finally annealed for 50 ps, all performed under an NVT thermostat with the LAMMPS code \cite{LAMMPS}. The atomic interactions during this process are characterized by a ReaxFF force-field for Hf-O systems \cite{Senftle2016}. The purpose behind this MD step is to generate a realistically randomized starting structure; to remove any remaining coordination defects and unrealistic bond lengths, the cell size and atomic positions of the annealed structure are relaxed using the cp2k DFT code \cite{Khne2020}. TiN electrodes are then attached along the transport direction. The interface distance between TiN and HfO$_2$ is optimized in cp2k to provide the lowest energy.

Next, oxygen vacancy ($V_O^{+2}$) and ion ($O^{-2}$) rearrangements under an applied voltage ($V_{app}$) are modelled with an in-house KMC code. KMC has been used previously to model physical processes governed by the harmonic transition state theory \cite{Henkelman2001}, and is capable of reproducing realistically stochastic behavior \cite{Andersen2019}. The inputs to this model are a selected set of events which can occur in the oxide, as well as the corresponding activation energies which would need to be overcome to execute them. Here we consider four types of events: (1) vacancy/ion pair generation, (2) vacancy/ion pair recombination, (3) vacancy diffusion (in which a vacancy exchanges lattice positions with an oxygen atom), and (4) oxygen ion diffusion (in which an oxygen ion moves between interstitial sites in the lattice). An event is then selected from this set once per timestep. The selection probability $P^x_{i,j}$ of event `x' occuring between locations (or `sites') `$i$' and `$j$' in the oxide is given by

\begin{equation}
    P^x_{i,j} = \nu \cdot exp\left(-\frac{E^x_A - E_{i,j}}{k_{B}T}\right),
\end{equation}

\noindent where $E^x_A$ is the zero-field activation energy of event x, and $\nu$ is an attempt frequency \cite{Zeumault2021}. The sites involved in each event are restricted to being within a predefined neighbor radius of one another. The zero-field activation energies $E^x_A$ are found for amorphous HfO$_2$ through DFT calculations, using the Nudged Elastic Band (NEB) method. We assume a reduced activation energy for vacancy generation at the interface with the active (top) electrode \cite{Traore2018}. Subtracted from this is the energy provided by the applied field $E_{i,j} = q\cdot(\upphi_i-\upphi_j)$, considering the charge ($q$) and potential ($\upphi$) for each pair of sites involved. Clustered vacancies are presumed conductive and first set to a charge state of zero ($V_O^{0}$). $\upphi$ at each site is then determined by treating the oxide as a network of nodes across which $V_{app}$ dissipates. 

The KMC algorithm makes a list whose elements $k_x$ correspond to the cumulative sum of event probabilities $P^{n}_{i,j}$, $n = 1 \cdots x$ and chooses the first element of the list that is greater than the sum of all event probabilities $k_N$, multiplied by a random number $r$, $r \in (0,1)$. The corresponding event $P^{x}_{i,j}$ is thus chosen to occur. The event selection-and-execution process repeats until the simulation timescale reaches the intended duration for which $V_{app}$ is applied. A device snapshot is finally generated, containing information on the atomic species and positions. \textbf{Figure 2} presents an overview of this model.

To calculate the conductance of each snapshot, we use the Quantum Transmitting Boundary Method (QTBM) as implemented in the OMEN code \cite{Luisier2006, Ducry2020}. Coherent transport occurs through states $\psi(E)$, which are found by solving: \begin{equation}
    (E\cdot S_{cp2k}-H_{cp2k}-\Sigma^{RB}(E))\cdot\psi(E)=Inj(E)
\end{equation}
\noindent Here, $Inj(E)$ describes carrier injection from the contacts, which are coupled to the device through $\Sigma^{RB}(E)$. $H_{cp2k}$ and $S_{cp2k}$ are the Hamiltonian and overlap matrices, respectively, as produced with cp2k for each of the KMC snapshots. Due to the localized nature of the underlying Gaussian-type orbitals, $H_{cp2k}$ and $S_{cp2k}$ have a block-structure representing atomic layers as pictured in \textbf{Fig. 1}. The $O^{-2}$ ions are not included in electronic structure calculations.

\begin{figure}
    \begin{overpic}[width=\columnwidth, tics=10]{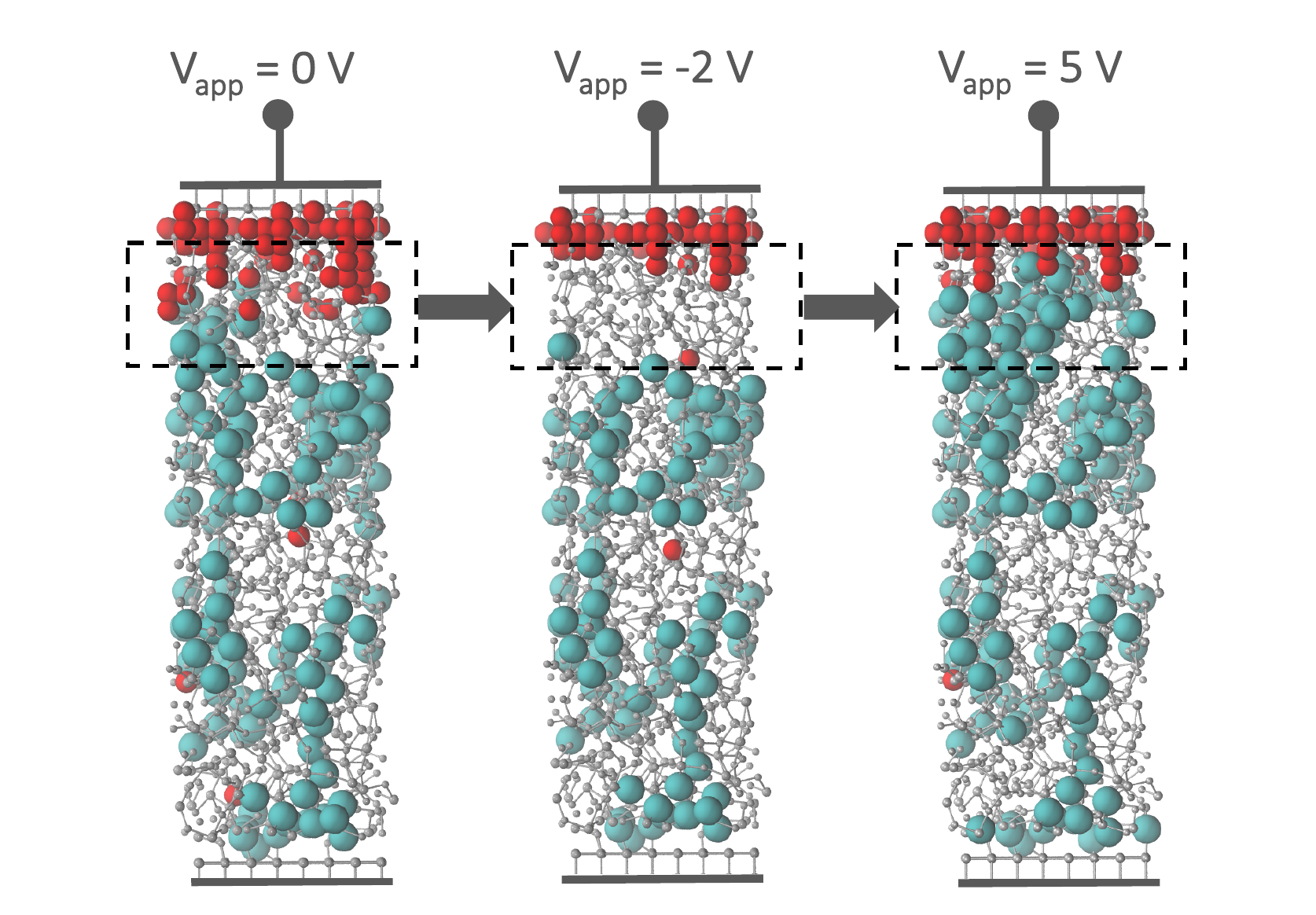}
    \put (5,65) {(a)}
    \put (35,65) {(b)}
    \put (63,65) {(c)}
    \end{overpic}
     \caption{Properties of the high (HRS) and low (LRS) resistance states of the TiN/HfO$_2$/TiN VCM cell during a single switching cycle. $V_O^{+2}$ and $O^{-2}$ positions along the oxide are shown for \textbf{(a)} the formed filament, \textbf{(b)} the HRS, and \textbf{(c)} the LRS. The areas near the active electrode, where the filament length varies most with the $V_{app}$, are indicated with dashed boxes.}
\end{figure}

\begin{figure}[h]
     \centering
     \begin{subfigure}[b]{0.49\columnwidth}
         \centering
         \begin{overpic}[width=\textwidth, tics=10]{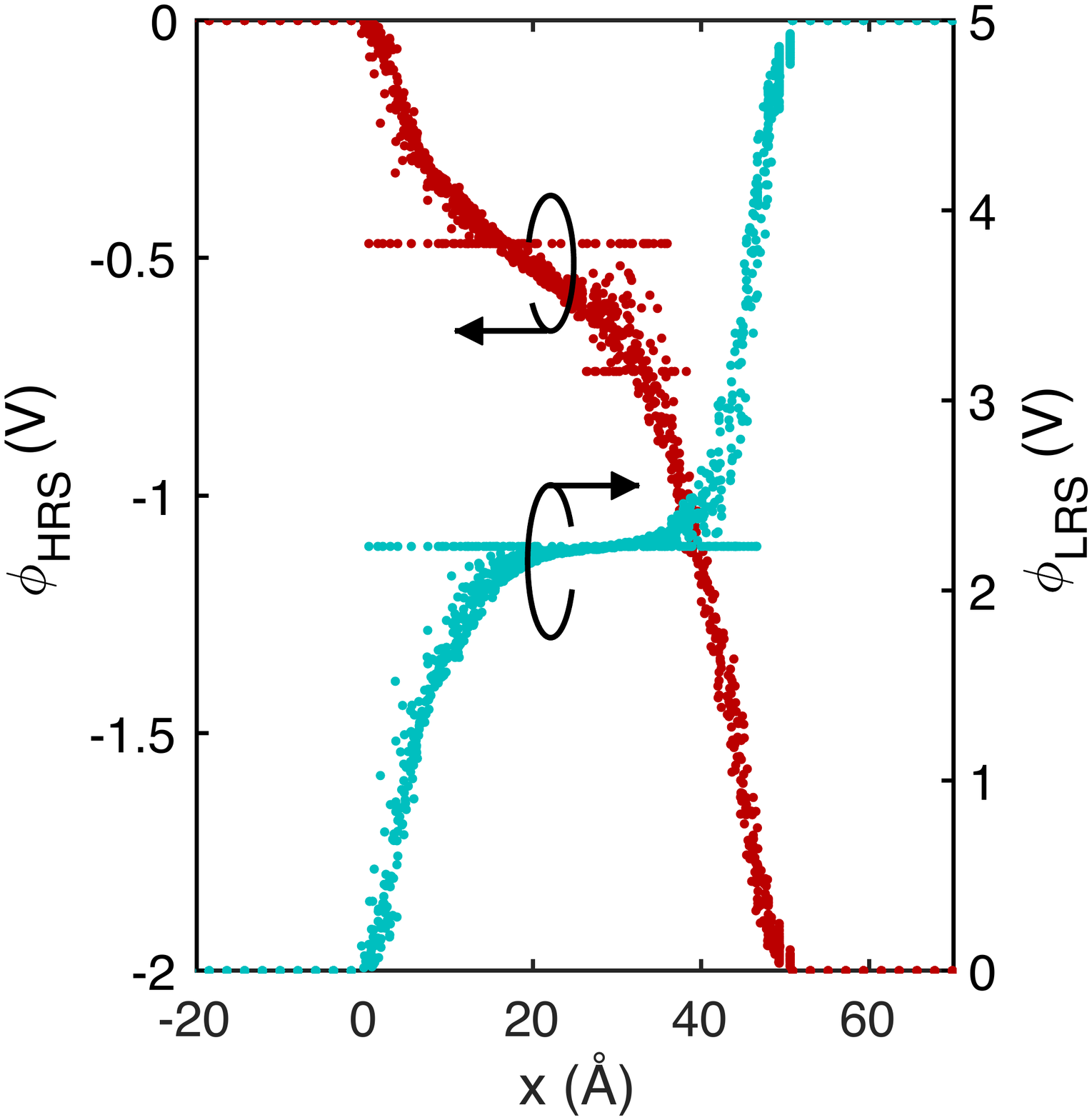}
         \put (0,100) {(a)}
        \end{overpic}
     \end{subfigure}
     \hfill
     \begin{subfigure}[b]{0.49\columnwidth}
         \centering
         \begin{overpic}[width=\textwidth, tics=10]{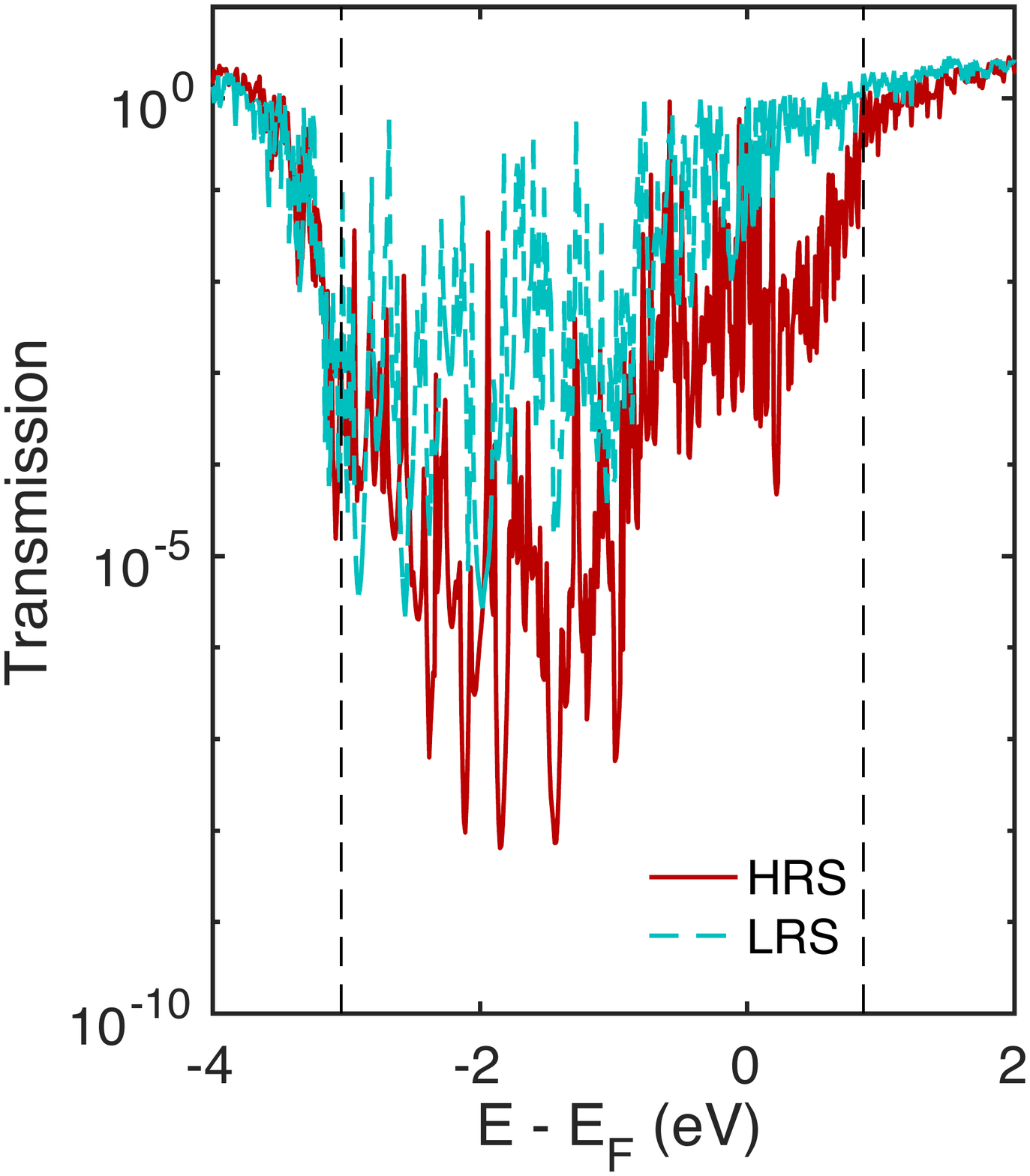}
         \put (0,100) {(b)}
        \end{overpic}
     \end{subfigure}
    %  \hfill
        \caption{Electrostatic potential ($\upphi$) along the oxide for the HRS (red) and LRS (blue) from \textbf{Fig. 3(b-c)}. The HfO$ _2$ oxide spans from 0 {\AA} to 50 {\AA}. \textbf{(c)} Transmission through the HRS and LRS states. The dashed lines indicate the estimated bandgap of the HfO$_2$ layer.}
\end{figure}

%***********************
% Results/Discussion
%***********************

\section{Results}

We apply this model to simulate the operation of a TiN/HfO$_2$/TiN VCM cell. The HfO$_2$ oxide has a cross-section of 1.67 x 1.63 $nm^2$, for a total of 2252 atoms. Starting from a structure with a formed filament (\textbf{Fig. 3a}), a $V_{app}$ of -2 V is applied to switch to the high resistance state (HRS, \textbf{Fig. 3b}). $V_{app}$ = 5 V is then applied to the HRS to recover a low resistance state (LRS, \textbf{Fig. 3c}). In each case, the duration of $V_{app}$ is 10 ms. 

\begin{figure}[t]
     \centering
     \begin{subfigure}[b]{0.49\columnwidth}
         \centering
         \begin{overpic}[width=\textwidth, tics=10]{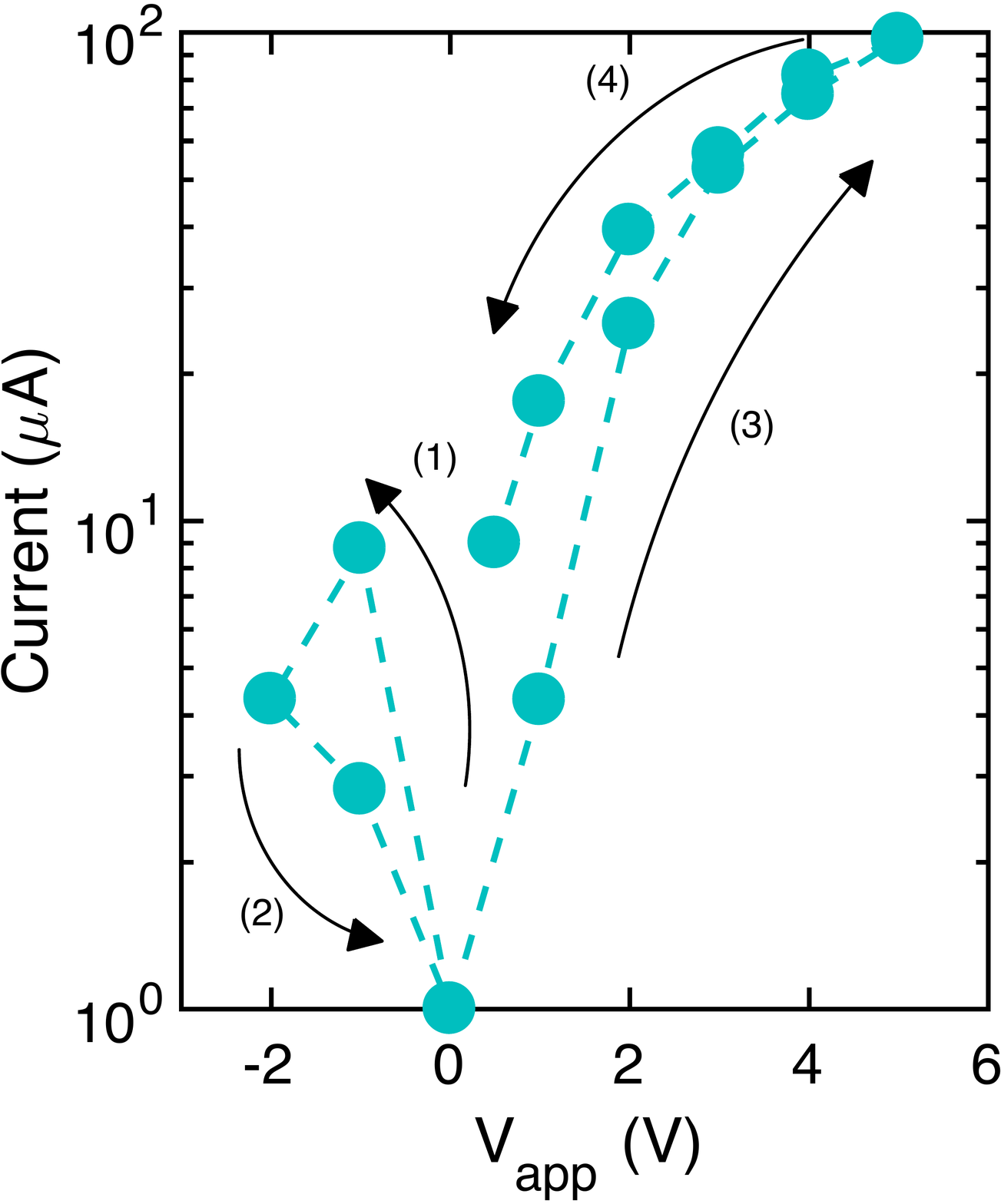}
         \put (0,103) {(a)}
        \end{overpic}
     \end{subfigure}
     \hfill
     \begin{subfigure}[b]{0.49\columnwidth}
         \centering
         \begin{overpic}[width=\textwidth, tics=10]{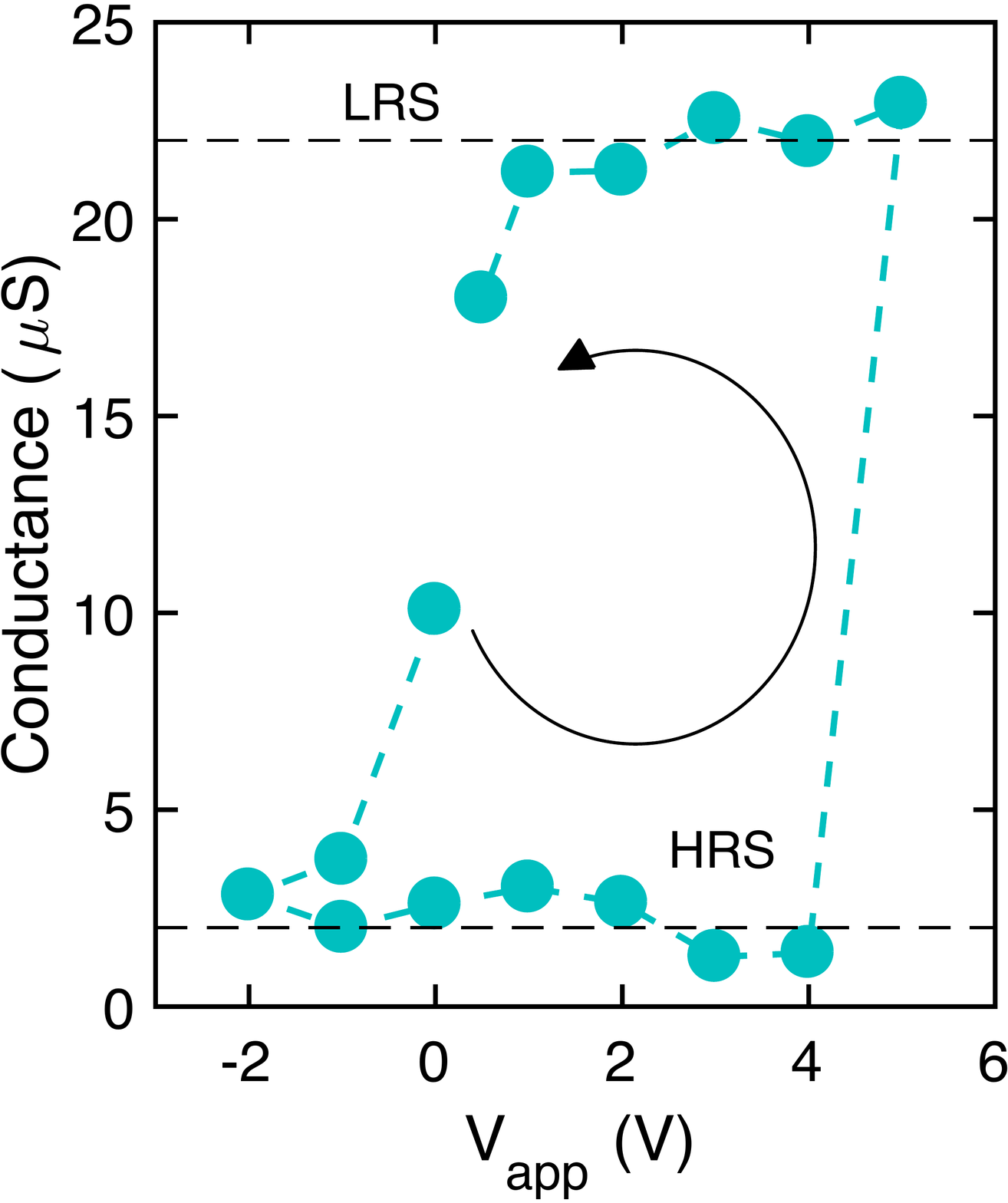}
         \put (-3,103) {(b)}
        \end{overpic}
     \end{subfigure}
        \caption{Current \textbf{(a)} and conductance \textbf{(b)} of the device shown in \textbf{Fig. 3} during a switching cycle, from the initial filament (0 V) to the HRS (-2 V) and finally to the LRS (5 V).}
\end{figure}

During the transition from the initial structure in \textbf{Fig. 3a} to the  HRS in \textbf{Fig. 3b}, recombination of oxygen ions with vacancies near the active electrode results in a local dissolution of the filament, creating a tunneling gap. A transition back to the LRS in \textbf{Fig. 3c} occurs when a sufficiently high $V_{app}$ regenerates vacancy/ion pairs in this gap. \textbf{Figure 4a} shows the potential for the HRS and LRS structures, as used during the KMC process. The steeper electric potential across the tunneling gap in the HRS increases the likelihood of vacancy regeneration in this area, assisting in the HRS-to-LRS transition. The difference in the transmission function between these two states is pictured in \textbf{Fig. 4b}, and is highest near the conduction band of HfO$_2$, consistent with the location of $V_O^{+2}$ vacancy defect states at this energy range \cite{Robertson2005}.

\textbf{Figure 5} plots a full switching cycle, showing both the current (\textbf{Fig. 5a}) and conductance (\textbf{Fig. 5b}) at each intermediate $V_{app}$. The device has a current hysteresis typical of VCM, with a conductance ratio of $\sim$10. In the HRS and LRS, the conductance values are relatively constant, indicating their non-volatility. The asymmetry in the $V_{app}$ between the HRS and the LRS transitions stems from the length of the conductive filament being lower in the initial device than in the final LRS, (see \textbf{Fig. 3a/c}), and from the recombination of $V_O^{+2}$/$O^{-2}$ being far more energetically favorable than their generation.

%***********************
% CONCLUSION
%***********************

\section{Conclusion}

We combined KMC and \textit{ab initio} quantum transport on an atomic lattice to model resistive switching in a VCM cell. Our model is able to capture a non-volatile switching mechanism dominated by vacancy generation and recombination at the active electrode, resulting in a clear conductance hysteresis with an ON/OFF ratio of one order of magnitude. This framework can be used to provide insight towards the optimization of VCM material stacks and geometries.

\section*{Acknowledgements}

We acknowledge funding from SNSF Sinergia (grant no. 198612), the Werner Siemens Stiftung Center for Single-Atom Electronics and Photonics, the Natural Sciences and Engineering Research Council of Canada (NSERC) Postgraduate Scholarship, as well as computational resources from the Swiss National Supercomputing Center (CSCS) under project 1119. 

\bibliography{main}

\end{document}